\documentclass[aip,rsi,reprint,graphicx]{revtex4-1} 
\usepackage{amssymb}
\usepackage{ulem}
\usepackage{float}
\usepackage{graphicx}
\usepackage{amsmath}
\usepackage{bm}
\usepackage{epsfig} 

\begin{document}

\title{Set-up of a High-Resolution 300 mK Atomic Force Microscope in an Ultra-High Vacuum Compatible $^3$He/10T Cryostat} 
\author{H. von Allw\"orden}
\author{K. Ruschmeier}
\author{A. K\"ohler}
\author{T. Eelbo}
\author{A. Schwarz}
\email[]{aschwarz@physnet.uni-hamburg.de}
\author{R. Wiesendanger}
\affiliation{Department of Physics, University of Hamburg, Jungiusstrasse 11, 20355 Hamburg, Germany}

\date{\today}

\begin{abstract}
The design of an atomic force microscope with an all-fiber interferometric detection scheme capable of atomic resolution at about 500 mK is presented. The microscope body is connected to a small pumped $^3$He reservoir with a base temperature of about 300 mK. The bakeable insert with the cooling stage can be moved from its measurement position inside the bore of a superconducting 10 T magnet into an ultra-high vacuum chamber, where tip and sample can be exchanged \textit{in-situ}. Moreover, single atoms or molecules can be evaporated onto a cold substrate located inside the microscope. Two side chambers are equipped with standard surface preparation and surface analysis tools. The performance of the microscope at low temperatures is demonstrated by resolving single Co atoms on Mn/W(110) and by showing atomic resolution on NaCl(001). 
\end{abstract}



\maketitle

\section{Introduction}

Atomic force microscopy (AFM) in the non-contact regime (NC-AFM) is widely used to image surfaces with atomic resolution in an ultra-high vacuum environment. In its spectroscopic modes of operation it is possible to map energy landscapes and dissipative processes on the atomic scale. Importantly, AFM can be applied to insulators as well, which distinguishes it from scanning tunneling microscopy (STM). A comprehensive overview on the different state of the art experimental methods and techniques can be found in the NC-AFM book series \cite{NCAFM1,NCAFM2,NCAFM3}. 

Although atomic resolution can be routinely obtained with room temperature set-ups, operating a force microscope at low temperatures is very beneficial, because (i) force sensitivity and energy resolution scale with $\sqrt{T}$; (ii) all thermally induced effects like thermal drift and adsorbate mobility are reduced; (iii) the atomic configuration at the tip apex is more stable allowing a controlled functionalization with, e.g., carbon monoxide molecules that can be used to record stunning submolecular resolution images \cite{LGross2009}. Last but not least, a number of very interesting physical phenomena only occur below a certain critical temperature, e.g., magnetic ordering or superconductivity. Therefore, many low temperature force microscopes have been realized in the past \cite{kirk88,giessibl91,albrecht92,hug93,yuan94,euler97,WAllers1998,pelekhov99,thomson99,rychen99,hug99,weitz00,volodin00,roseman00,suehira01,MLiebmann2002,MHeyde2004,TAn2008,BJAlbers2008,UGysin2011}. Some sub-Kelvin AFM-type set-ups dedicated to rather special experiments, like magnetic resonance force microscopy \cite{HJMamin2001}, have been presented in the past. However, low temperature UHV AFM set-ups capable of atomic resolution were limited to $^4$He cryostat systems. 

Here, we present the design of a bakeable ultra-high vacuum (UHV) system with an atomic force microscope attached to a $^3$He reservoir of a commercially available $^3$He cryostat \cite{OI}. The microscope features a Rugar type all-fiber interferometric detection scheme \cite{DRugar1989} and is located inside the bore of a 10 T superconducting magnet. Its main purpose is the investigation and - utilizing the magnet - manipulation of single magnetic atoms and molecules on insulating substrates using magnetic exchange force microscopy \cite{MExFM} and spectroscopy \cite{MExFS}.

Note, that we decided for an optical detection scheme and not for the more recently developed qPlus design \cite{FGiessibl1998}. The latter has a very low power consumption and provides excellent atomic resolution as well \cite{NCAFM2,NCAFM3}, but unlike for STM set-ups the preamplifier needs to be nearby the tip to avoid excessive noise. Thus, to avoid charge carrier freeze out, the preamplifier must be heated, whereby the advantage of low power consumption is lost. Even worse, large magnetic fields required for our envisaged experiments influence the motion of charge carriers in the semiconducting elements and hence perturb the function and operation of the preamplifier.  

Currently, the base temperature of the microscope body under measurement conditions (laser on) is 540 mK with a hold-time of about 7 hours. We demonstrate that our set-up is able to achieve atomic resolution on bulk insulators like NaCl(001) and that clean metallic samples like Mn/W(110) with single Co atoms deposited in-situ onto the cold surface can be prepared and investigated. Finally, we discuss improvements to lower the base temperature and increase the hold time.

\begin{figure*}
\centerline{\includegraphics[width=0.8\textwidth]{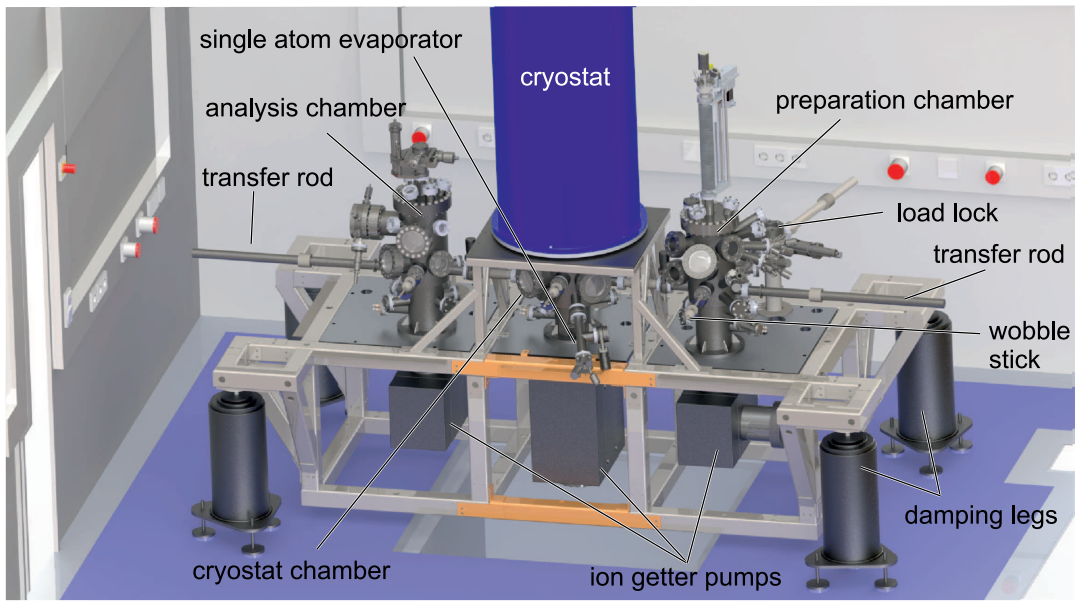}}
\caption{Layout of the UHV-system in the lower floor of the two-story lab. Three chambers are aligned symmetrically in a row with the cryostat placed on top of the central vessel. The frame that carries the UHV-system is supported by four passive damping legs. The lab has a separate foundation. All electronic devices required to operate the microscope are located in the upper floor of the lab.}
\label{fig1}
\end{figure*}

\section{UHV System}

The UHV system is comprised of a load lock, a preparation chamber, an analysis chamber and a central chamber with the cryostat on top. Figure \ref{fig1} displays the layout of the system located in the lower floor of a two-story lab. All electronics to operate the microscope are located in the upper floor.

The whole system is mounted to a frame made of hollow stainless steel bars that are filled with sand to damp acoustic waves. The frame itself is placed on four passive vibration isolation damping legs, which stand on a concrete foundation separated from the rest of the building. A rigid bake-out tent with heaters, fans and a temperature control can be attached to the frame. Analysis and preparation chamber are symmetrically placed to the sides of the cryostat chamber to reduce the number of eigenmodes of the whole construction and to ensure that the center of mass is as close as possible to the geometric center. Tip and sample transfer between the different chambers is performed with magnetically coupled manipulators. Specific locations for tip and sample within a chamber, e.g., the heating stage, can be accessed via wobble sticks. All chambers are pumped with scroll pumps, turbomolecular pumps and vibrationless ion getter pumps (only the latter are in operation during measurements) to achieve a base pressure of about $1\times 10^{-10}$~hPa after bake-out for a few days at $120^\circ$C.

\subsection*{Load Lock and Preparation Chamber}

Samples and tips are introduced into the system via a small bakeable load lock chamber. It features a cleavage station with integrated heating stage suitable to prepare ionic crystals like NaCl.
 
For further sample treatment the preparation chamber features a home-built vertical manipulator in the center of the preparation chamber with a resistive heater, a quartz crystal micro-balance (QCMB) and a tungsten hook. The sample can be hung into the hook using a wobble stick. A moveable filament can be approached to the backside of the sample. This e-beam heating stage can generate temperatures well above $2000^\circ$C by applying up to 2000 V between sample and filament. The required power to reach a particular sample temperature depends on the distance between filament and sample. For a typical spacing of 2 mm about 125 W are needed to reach $1900^\circ$C, which can be measured with a pyrometer from the outside through a window. Note that separations being too small could bend the sample holder towards the filament due to strong electrostatic attraction and in the worst case fuse them together. Since the sample is only attached to a hook far away from any other components, heating of the surrounding is minimal. As a result, the background pressure during high-temperature cleaning of crystals like tungsten remains in the $10^{-10}$~hPa regime.

If lower temperatures are needed ($<800^\circ$C), e.g., for annealing, the sample can be placed onto a resistive heater plate. In this case the temperature can be simply monitored with a thermocouple. By moving the sample vertically and rotating it towards different ports, cleaning by argon ion bombardment or deposition of various metals from e-beam evaporators with integrated flux monitors is possible. If needed, the sample can be kept at elevated temperatures during argon ion bombardment or metal deposition. To independently check the evaporation rate, the QCMB can be moved in front of the evaporators instead of a sample. Gases like oxygen (for cleaning by oxygen glowing) or argon (for cleaning by ion bombardment) are admitted via leak valves attached to the chamber.

\subsection*{Cryostat and Analysis Chamber}

In the cryostat chamber the \textit{in-situ} tip and sample exchange can be carried out with a wobble stick. Moreover, single atoms can be deposited from an e-beam evaporator while the substrate is still cold (about 25 - 30 K). To do so, the insert with the microscope has to be moved from the upper measurement position (in the center of the magnet) to the lower transfer position (in the center of the cryostat chamber). To provide sufficient visibility and illumination several large and small viewports are attached to the chamber. In particular, two side viewports in combination with two holes inside the microscope allow quick pre-adjustment of sample and cantilever. Note that due to the thermal contraction the final adjustment of the fiber and sample approach has to be done in the upper measurement position. 

The analysis chamber is equipped with low energy electron diffraction (LEED) and Auger electron spectroscopy (AES) unit for characterizing substrate quality before performing low temperature experiments. It also has a storage facility large enough to carry 24 tips or samples. Spare flanges provide ample possibilities to attach additional equipment. To transfer tips and samples between preparation and analysis chamber the wobble stick in the cryostat chamber is used to perform a handover between the transfer rods attached to both side chambers.

\section{Cryostat}
 
\begin{figure}
\centerline{\includegraphics[width=0.9\linewidth]{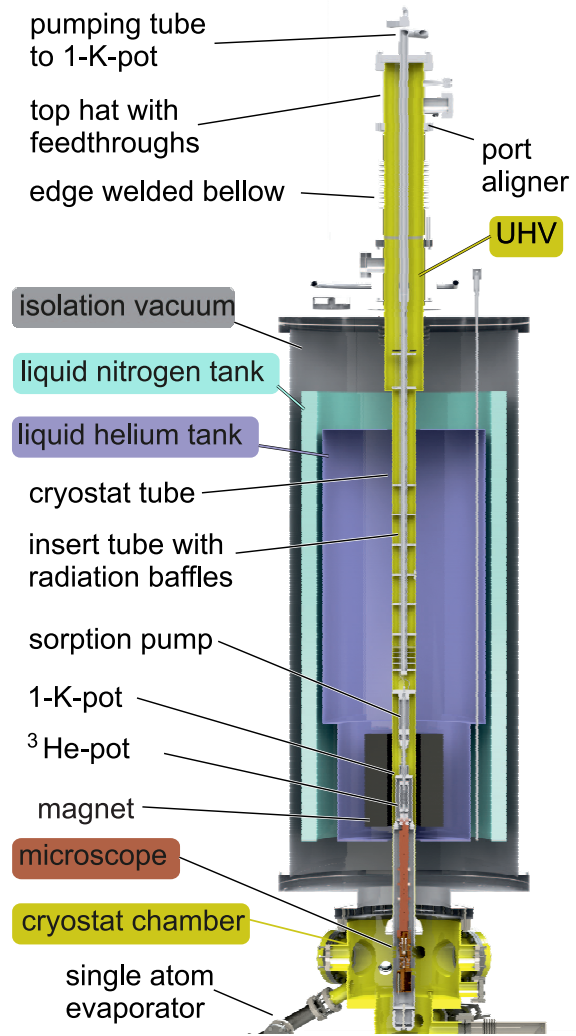}}
\caption{Sectional view of cryostat and cryostat chamber. The outer section of the cryostat contains the isolation vacuum, which is separated from the UHV-system, and the liquid nitrogen as well as the liquid helium reservoir. The cryostat tube connects top hat and cryostat chamber (yellow). It contains the moveable insert with the 1-K-pot, the sorption pump and the $^3$He-pot with the microscope body attached to it. The movement of the insert from the upper measurement position in the center of the magnet to the lower position in the center of the cryostat chamber is facilitated by a motor-driven edge welded bellow mounted on top of the cryostat.}
\label{fig2}
\end{figure}

The cryostat is based on a commercial UHV-compatible Heliox system from Oxford Instruments \cite{OI} using the $^3$He evaporation principle \cite{Kugler2000,Wiebe2004}, but with an edge welded bellow instead of a spindle mechanism to move the insert. A cross sectional view is depicted in Fig. \ref{fig2}. The cryostat tube is located along the symmetry axis of a cylindrical evacuated Dewar with an outer liquid nitrogen tank (capacity: 75 l, hold time: 10 days) and the main liquid $^4$He tank (capacity: 100 l, hold time: 6 days). Filling cryogenic liquids via siphons is done from the upper floor of the lab. A superconducting magnet made from NbTi and Nb$_{3}$Sn producing a field of up to 10 T perpendicular to the sample surface is located inside the $^4$He tank. A 90 mm diameter bore provides sufficient space for the cryostat tube. The tube itself is multi-walled, so that hot dry nitrogen can be circulated to heat up the innermost wall while keeping the outermost wall, which is close the superconducting magnet, as cold as possible during bake-out. Nevertheless, to protect the magnet, it has to be immersed in liquid nitrogen during bake-out. For low temperature operation a needle valve allows to fill the multi-walled tube with $^4$He from the main tank. 

The lower end of the cryostat tube is connected to the cryostat chamber while the upper end is connected to an edge welded bellow with a small chamber on top. The insert with the $^3$He unit described in the following section is connected to the top flange of this small chamber. All electrical feedthroughs and the feedthrough for the glass fiber required to operate the microscope as well as the $^3$He unit are attached to this chamber. To prevent room temperature radiation from entering the cryostat tube, its bottom is closed by spring loaded flaps, which open when the microscope is moved downwards. To increase the thermal path from the $^4$He bath and the liquid nitrogen tank to the outer walls at room temperature and to compensate for any differential thermal expansion between 4 K (low temperature operation) and 400 K (bake-out temperature), a hydroformed bellow assembly is implemented. Only UHV compatible, i.e., bakeable, CF flanges with copper gaskets (instead of, e.g., indium gaskets) were employed. 

The UHV region extents from the top hat along the edge welded bellow and the cryostat tube down to the cryostat chamber below (yellow area in Fig. \ref{fig2}). Additional non-evaporable getter pumps (NEG pumps) are attached to the top-hat und just below the edge welded bellow. They pump hydrogen very well and thus reduce the amount of hydrogen cryopumped by the cold walls of the cryostat tube. This in turn minimizes the release of hydrogen during movement of the insert due to sliding friction and thereby prevents contamination of the sample surface.

\subsection*{Insert}

The insert consists of a stainless steel pumping tube attached to the top flange mounted onto the small top chamber and connected to a 1-K-pot and the  $^3$He-pot. The former can be pumped by an external rotary vane pump attached to the pumping tube, while the latter is pumped by a vibrationless charcoal sorption pump that is an integral part of the insert. The microscope is attached via a slit copper rod to the $^3$He-pot and surrounded by a cylindrical 1-K-radiation-shield connected to the bottom of the 1-K-pot. To block room temperature radiation from the top chamber several metallic discs (radiation baffles) with a diameter slightly smaller than the inner diameter of the insert tube are attached to the pumping tube. All electrical wires as well as the optical fiber are thermally anchored along the pumping tube, at the sorption pump that is directly in contact with the 1-K-pot, and at the rod that connects microscope body and $^3$He-pot.

To replenish the 1-K-pot with liquid Helium from the main tank, a flexible stainless steel capillary spiraling between the 1-K-shield and the inner wall of the cryostat is used. The capillary can be detached by a VCR Swagelok connection. The filling rate can be adjusted via a needle valve. By pumping on the 1-K pot, the temperature of the $^4$He can be reduced below its boiling temperature of 4.2 K. This is required to liquefy $^3$He that has a boiling temperature of about 3.2 K. 

If not in use, the $^3$He is stored in a room temperature dump outside the cryostat. The dump is equipped with a valve that can be opened to admit $^3$He via a thin straight capillary to the $^3$He-pot and the sorption pump. Both are linked via a 60 mm long hydroformed bellow with a diameter of about 10 mm that also opens and closes a thermal switch. The thermal switch is composed of the conical end at the bottom of the $^3$He-pot and a counter cone connected to the 1-K-radiation-shield. If the $^3$He is gaseous, the bellow expands and presses the $^3$He-pot into the countercone establishing a direct thermal contact to the 1-K-pot.

\subsection*{Cool Down Procedure}

Initial cool down requires that nitrogen and the main $^4$He-tank are filled with liquid nitrogen and helium, respectively. Additionally, the two needle valves between the main $^4$He-tank and the multi-walled cryostat tube as well as the 1-K-pot need to be open. In this condition the insert is cooled via radiation to the cryostat tube wall and via the 1-K-pot. However, since 1-K-pot and $^3$He-pot are thermally isolated from each other, cool down would take very long.

To drastically increase the cooling rate, $^3$He is utilized as exchange gas. Therefore, the valve between $^3$He dump and $^3$He section of the insert is opened. $^3$He diffuses into the cold sorption pump and the $^3$He-pot. If the sorption pump is colder than 20 K, $^3$He is retained in the charcoal. If all $^3$He is adsorbed, the valve to the dump at room temperature is closed. Subsequently, the sorption pump is heated to about 40 K, whereby all $^3$He is released and can act as an exchange gas. Additionally, the large pressure of the gaseous $^3$He closes the thermal switch, which accelerates the cool down further. After about one day the insert is cooled down below 10 K. 

To reach base temperature the $^3$He has to be liquefied. To do so, the 1-K-pot is pumped, whereby its temperature is reduced down to about 2 K while the sorption pump is still heated to 40 K. As a result, the $^3$He, which condenses at about 3.2 K, liquefies and is collected in the $^3$He-pot. If all $^3$He is liquefied, the pressure decreases and the thermal switch is open again, whereby the $^3$He pot is thermally isolated from the 1-K-pot. Now the heater for the sorption pump is switched off. Below about 20 K the sorption pump effectively pumps on the $^3$He-pot, whereby its temperature is reduced below 3.2 K. Since the 1-K-pot is constantly replenished via the capillary from the main $^4$He tank, the hold time in this so called continuous mode of operation is not limited by the size of the 1-K-pot, but by the total heat load and the amount of liquid $^3$He; cf. table \ref{tab1}. The lowest achievable temperature also depends on the pumping speed of the sorption pump.

To reduce the vibrational noise during measurements the 1-K-pot pump can be switched off. In this so called single shot mode of operation the temperature in the 1-K-pot increases from about 2 K to 4.2 K, whereby the sorption pump temperature increases as well, which reduces its pumping speed slightly. Furthermore, the thermal load increases as well due to an increased thermal radiation from the 1-K-radiation shield. As a result, the final temperature increases and the hold time decreases compared to the continuous mode of operation; cf. table \ref{tab1}. Note that in this mode another source of vibrational noise is eliminated as well: In the continuous modes $^4$He from the main tank enters the suprafluid $^4$He in the pumped 1-K-pot ($^4$He becomes suprafluid below 2.2 K). The mixing turbolences create vibrations \cite{PGorla2004}. If $^4$He in the 1-K-pot is also in its normal phase, like in the single shot mode, this noise is of course absent.

\begin{table}
\centering

\begin{tabular}{l|cc}

condition & \quad $t_{\rm h}$  & \quad $T_{\rm b}$ \\

\hline

\rule{0cm}{0.5cm}
$^3$He-pot without microscope, & 42 h & 307 mK \\
continuous operation& & \\
$^3$He-pot without microscope, & 21 h & 333 mK \\
single shot operation& & \\
$^3$He-pot with microscope, & 12 h & 350 mK \\
single shot operation, laser off & & \\
$^3$He-pot with microscope, & 7 h & 500 mK \\
single shot operation, laser on & & \\
\end{tabular}

\caption{Hold times $t_{\rm h}$ and base temperatures $T_{\rm b}$ for different conditions. Continuous (single shot) operation means that the 1-K-pot is pumped (not pumped). AFM experiments can only be performed without pumping the 1-K-Pot and of course with laser on (last row). Here, $T_{\rm b}$ and $t_{\rm h}$ are limited by stray light, which can be reduced by mirror coating fiber end as well as cantilever backside.}

\label{tab1}

\end{table}

\section{Microscope Design}

The design of the microscopy body itself is similar to our previously built instruments \cite{WAllers1998,MLiebmann2002}, but made from phosphorous bronze instead of MACOR (because of its larger thermal conductivity) and with a smaller diameter (because of space constraints within the bore of the magnet). As mentioned in the introduction we choose optical detection, because the performance of the employed Rugar-type all-fiber interferometer  \cite{DRugar1989} is not compromised by the presence of strong magnetic fields. Moreover, commercially available micro-machined beam-type cantilevers provide a much wider range of spring constants and resonance frequencies and thus allow to perform a larger variety of experiments than available qPlus sensors. In the following, we explain details of our design concept that are specific to this $^3$He system. More general design criteria can be found in descriptions of our earlier 4-K-instruments \cite{WAllers1998,MLiebmann2002}.

Since the microscope is operated in a vacuum system, a reliable remote controlled mechanism to approach the sample to the tip and the optical fiber to the cantilever backside has to be implemented. As in previous designs \cite{WAllers1998,MLiebmann2002}, we used Pan-type piezo-electrically driven stepper motors with six legs. However, to reduce the thermal load from wires, we selected the stick-slip configuration (one wire plus ground for each stepper motor) and not the walker configuration (six wires for each plus ground for each stepper motor). The cylindrical microscopy body is made from one piece of phosphorous bronze with a diameter of 35 mm and a length of 120 mm. Some smaller parts like the leaf spring for the stepper motor are fixed with screws to the body, while for others, like the piezo legs of the stepper motor, epoxy glue is used. Electrical connections are either made by soldering or using conductive glue. To facilitate fast thermalization the body is made from phosphorous bronze which possesses a much higher heat conductivity (0.2 W/(m$\cdot$K) at 1 K \cite{Lakeshore}) compared to MACOR (0.004 W/(m$\cdot$K) at 1 K \cite{Runyan2008}). However, it can be expected that the differential thermal expansion between piezo material PZT and phosphorous bronze is much larger than for MACOR \cite{GNunes1995}. We found that direct gluing is fine for small contact areas (a few mm$^2$), while for larger areas screwed ceramic spacer (MACOR or Al$_2$O$_3$) should be employed.

\begin{figure}
\centerline{\includegraphics[width= 0.9\linewidth]{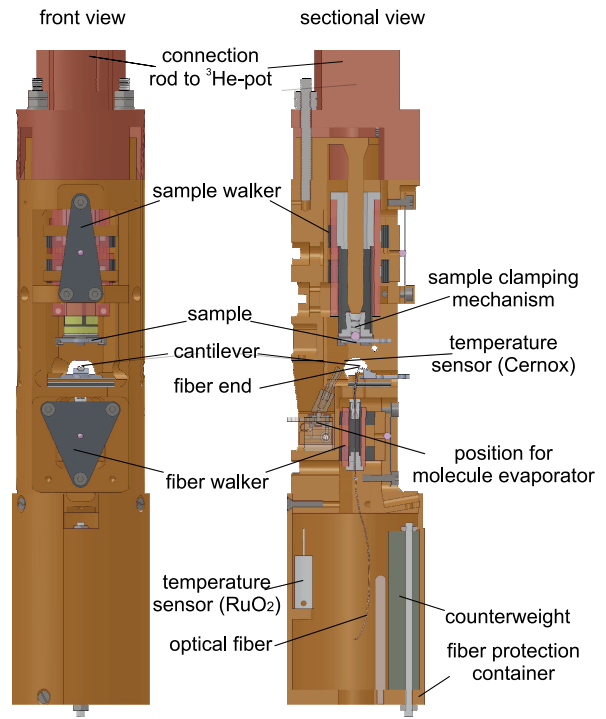}}
\caption{Front and cross sectional view of the microscope body. The fixed cantilever stage is located in the middle. The fiber and the sample can be approached to the cantilever backside and to the tip attached to the free end of the cantilever, respectively, using two piezo-driven stepper motors denoted as fiber and sample walker.}
\label{fig3}
\end{figure}

A CAD-model of the microscope is displayed in Fig. \ref{fig3}. In the central part the cantilever stage with the shaker piezo is visible. It consists of two phosphorous bronze brackets screwed onto a MACOR plate with a wear-resistive Al$_2$O$_3$ plate in between. The latter prevents abrasion due to repetitive cantilever exchanges. The exchangeable cantilever holder is designed to form two arms which act as a spring. They are pressed together by the two brackets, if pushed into the stage. The shaker piezo, which is needed to apply the energy for the cantilever excitation, is sandwiched between two MACOR plates for electrical isolation. The lower one is screwed onto the microscope body. Instead of a cantilever, an STM tip can be placed on the holder as well. Leads are provided to connect a current preamplifier outside the top vacuum chamber to perform STM measurements.

Detection of the cantilever position is realized via a Rugar-type interferometer set-up \cite{DRugar1989}. However, in our case the fiber end is coated with a dielectric layer that increases the reflectivity for our laser wavelength of 780 nm from about 4\% (reflectivity at the glass-vacuum interface) to about 25\%. Since this corresponds roughly to the expected reflectivity from the backside of a Si cantilever, the fringe visibility increases. The fiber is brought into UHV from the top and has to be turned by $180^\circ$ to approach the cantilever from the bottom. To safeguard the fiber and to store additional length a fiber protection container is attached to the bottom end of the microscope body. The fiber end is guided through a piezo-electric tube scanner and glued to its end. The fiber tube scanner is used for fine adjustment of the vertical distance between fiber end and cantilever backside. The tube itself is glued into the central bore of a triangular sapphire prism. Coarse vertical positioning of the fiber end relative to the cantilever backside is realized by a piezoelectrically driven stepper motor, which consists of six piezo legs that are in contact with the outer faces of the sapphire prism.

The sample stage is mounted on the piezo-electric xyz-tube scanner facing downwards. Its lateral scan size and maximum z-stroke for 220 V is $3.3 \mu$m$^2$ and 550 nm at 300 K, $1.6 \mu$m$^2$ at 275 nm for 14 K, and $0.8 \mu $m$^2$ and 95 nm at 500 mK. Clamping of the sample holder is facilitated by a spring loaded ruby ball inside the stage, which presses the sample holder backside against three lobes of the stage, which form a well-defined three point pressed contact. The tube scanner is glued into the bore of a hexagonal sapphire prism. Three of the outer faces of the sapphire are in contact with the piezo legs of the stepper motor. In this manner the sample can be approached to and retracted from the tip. 

The step size of the two stepper motors depends on the applied voltage as well as on the applied load between piezo legs and prism, which can be adjusted via a leaf spring. The spring must be carefully adjusted in ambient conditions to make sure that the motor still moves in UHV and at low temperatures while providing sufficient rigidity to allow for atomic resolution imaging. To exchange cantilever and sample holders, a wobble stick is used. An additional wobble stick is located at the back of the microscope to fix the microscope body during such an exchange or to insert a molecule evaporator directly into a small receptacle located at the backside of the microscope. The miniaturized evaporator \cite{KLaemmle2010} is simply a tiny crucible mounted onto a holder that has the same size as the sample and cantilever holders. It is made in a way that upon sliding the holder into the receptacle two electrical contacts are established that allow heating the crucible. This design allows evaporating molecules onto the cold substrate without lowering the microscope into the cryostat chamber.

The microscope body is not directly attached to the $^3$He-pot, but connected via a 315 mm long Cu rod with a diameter of 35 mm. This rod serves two purposes: First, it provides more surface area to thermally anchor the electrical wires to the $^3$He-pot. Moreover, the wires can be disconnected via IC-pin connectors to make the assembling and disassembling easier. Second, if the microscope is moved into the transfer position, $^3$He-pot, sorption pump and 1-K-pot are still in the cryostat tube and thus surrounded by 4 K. Thus, the microscope is still cooled and only warms up to about 30 K, which allows deposition of single atoms onto a cold substrate to prevent aggregation to larger clusters.

The wires, which are required to operate the low temperature section of the cryostat such as the temperature sensors and heaters of the $^3$He-pot, were provided by the manufacturer. All other wires necessary to operate the microscope were installed later. Different wires were chosen for the cabling of the microscope on the basis of mainly two conflicting criteria. First, the heat load should be as small as possible. Second, delicate measurement and control signals, such as those of the piezo tube scanners, had to be effectively decoupled and shielded from external noise sources. This was realized by choosing either shielded coaxial or shielded twisted pair cables. 

In order to be able to effectively cool those wires, which exhibit multiple insulation and conducting parts, they are thermally anchored to the insert at different positions. The contact length of the shielded wires was maximized by winding them around the cylindrical housing of the sorption pump that is at about 5 K during normal operation.  The wires were not thermally anchored to the 1-K-pot, which is usually at a similar temperature as the sorption pump (particularly in single shot mode, if the 1-K-pot pump is not running). The thermal resistance of the wires was improved by increasing the length of the wires between the $^3$He-pot (and thus the microscope) and the sorption pump. This was realized by loosely winding the wires around the rod which connects the microscope and the $^3$He-pot before establishing a tight contact by gluing them to the rod. In the end the effective length of the wires between the sorption pump and the microscope adds up to at least 500 mm. 

\begin{figure}
\centerline{\includegraphics[width=1.0\linewidth]{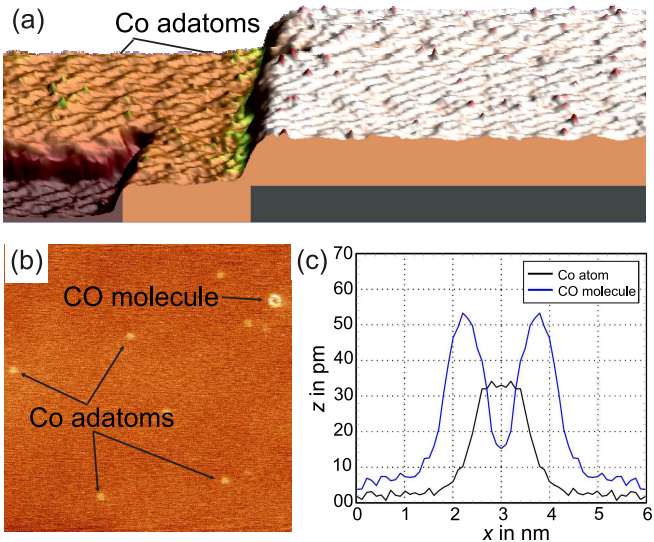}}
\caption{(a) Perspective view of a 250 nm $\times$ 86 nm area of a W(110) substrate covered with a submonolayer of Mn (about 55\%), onto which a low dose of Co was evaporated. Vibrational noise is visible as a modulation on top of the topography signal. Nevertheless, Co adatoms are clearly detectable as separated small protrusions. (b) 50 nm $\times$ 50 nm topography of a Mn covered area showing several Co adatoms with only one contamination feature that can be easily identified as CO molecule: While Co-adatoms appear as simple protrusions, CO-molecules exhibit a donut-like shape, as depicted by the line sections displayed in (c) \cite{ASchwarz2014}. Measurements were performed at 5.2 K.}
\label{fig4}
\end{figure}

\section{Thermometry}

Generally, calibrated Cernox and RuO$_2$ temperature sensors, both from Lake Shore \cite{Lakeshore} were used. The former are reliable from about 1.2 K to 425 K, i.e., also during bake-out, while the latter can be used to measure temperatures accurately down to the 100 mK range and below using an ac-bridge to reduce self-heating. Therefore, one Cernox and one RuO$_2$ temperature sensor were fixed to the microscope body as well as to the $^3$He-pot. Additionally, one Cernox sensor was attached to the 1-K-pot and one to the sorption pump. With this set of sensors it is possible to completely monitor the temperature of the insert during bake-out and cool down. They are also used to control heaters attached to the $^3$He-pot as well as to the 1-K-pot.
  
Conceptually, a force microscope possesses an intrinsic temperature sensor, i.e., the cantilever. By measuring the thermal spectral noise density of the cantilever eigenmode and applying the equipartition theorem to the Lorentzian profile it is possible to determine the cantilever temperature. However, in general only the temperature of the cantilever eigenmode is determined and due to photothermal forces and radiation pressure, the eigenmode can be cooled or heated relative to the equilibrium temperature of the surrounding \cite{CMetzger2004,HHoelscher2009}. A detailed analysis of these effects, including the effect of laser intensity noise, on the cantilever dynamics and  formulas to deduce the thermodynamic temperature of the cantilever by measuring its mode temperature can be found in Ref. \cite{GFlaeschner2015}. We find that, if the temperature of the microscope body is about 500 mK, the thermodynamic cantilever temperature is typically about 5 K due to light absorption of the Si cantilever. To reduce the cantilever temperature, its backside could be coated with a high-reflective metallic or dielectric mirror.

\section{Experimental Results}

Test measurements were conducted to characterize the performance of the microscope. As a metallic test sample a submonolayer Mn on W(110) was prepared. This is a demanding sample system, because it is prone to contaminations, if the pressure during preparation or transfer is too high. On the other hand, it is a magnetically interesting sample, as a monolayer Mn on W(110) exhibits an out of plane antiferromagnetic spin spiral in the [1$\bar1$0] direction with a periodicity of about 12 nm and a relative spin orientation of adjacent Mn atoms of $173^\circ$ \cite{Bode1}. Therefore, this sample system is a promising candidate for MExFM measurements itself and as a substrate for  single magnetic atoms or molecules. The W(110) crystal was cleaned by repeated cycles of oxygen glowing at about 1500 K for 30 minutes and flashing at a temperature of 2200 K for 15 s \cite{MBode2007}. Then Mn was evaporated from a crucible onto the W(110) at a constant rate as checked by a QCMB before and after the evaporation procedure. The temperature of the crystal was kept constant at $110^\circ$C during evaporation and the following annealing process, which lasted 10 min. The pressure of the system during Mn evaporation did not exceed $2.9 \cdot 10^{-10}$ hPa. Afterwards, the sample was transferred to the cryostat chamber and inserted into the microscope in order to check the quality of the sample preparation. Subsequently, the microscope was lowered into the cryostat chamber and Co atoms were evaporated from an electron beam evaporator attached to the cryostat chamber at a flux current of I$_{\rm flux }= 10$ nA for about 3 s onto the cold Mn/W(110) sample inside the microscope.

The result of such a preparation is displayed in Fig. \ref{fig4}(a): Parts of the W(110) substrate are covered with one atomic layer of pseudomorphically grown Mn. Small protrusions indicate the positions of individual Co adatoms. The area imaged in (b) shows eight well separated Co atoms in an area of 50 nm $\times$ 50 nm and one contamination with a different appearance. Deposition of additional Co atoms for about 8 s using the same parameters as before led to a nearly threefold increase of the number of protrusions. Thus, protrusions can be identified as single Co atoms. The donut shaped adsorbate visible in the upper right corner of Fig. \ref{fig4}(b) can be identified as upright standing CO molecule. This peculiar contrast feature originates from a local electrostatic repulsive interaction between the electrostatic dipole moments at tip apex and molecule, both pointing with their positive pole towards each other \cite{ASchwarz2014}. Note that the lateral size of an adsorbate displayed in (c) is related to the finite size of the tip apex. Also, the apparent height of the adsorbate is not the true geometrical height, but depends on the tip-sample distance as well as on the sharpness of the tip (sharper tips lead to larger apparent heights, because the long-range forces from the substrate are less dominant, if the tip is located above an adsorbate).

\begin{figure}
\centerline{\includegraphics[width=1.0\linewidth]{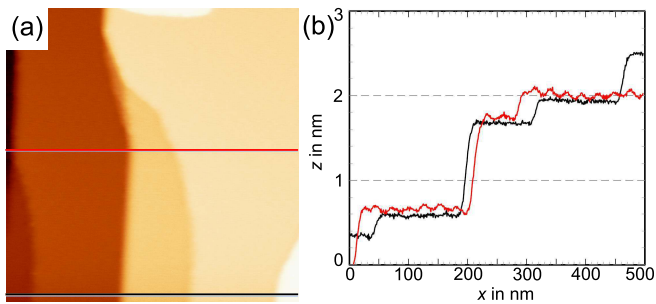}}
\caption{(a) 500 nm $\times$ 500 nm topography of a cleaved NaCl(001) sample recorded at 519 mK. Terraces with a width of up to several hundred nm are separated by step edges with a height of one or more interlayer distances $a/2$, as shown in the line section displayed in (b).}
\label{fig5}
\end{figure}

\begin{figure}[b]
\centerline{\includegraphics[width=1.0\linewidth]{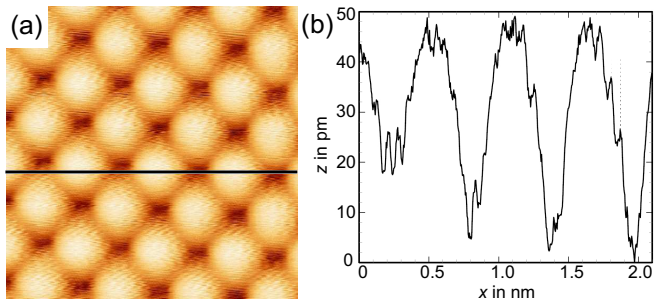}}
\caption{(a) Atomically resolved NaCl(001) of an area of 2 nm $\times$ 2 nm recorded at 548 mK. (b) The line section clearly shows the atomic corrugation as well as vibrational noise on top of it. The atomic corrugation of about 60 pm is visible with a peak-to-peak noise of about $\pm 4$ pm.}
\label{fig6}
\end{figure}

The second system was NaCl(001), prepared by \textit{in-situ} cleavage. It is an important test system, because only atomic force microscopy allows investigating bulk insulators with atomic resolution. NaCl(001) is a well-known ionic crystal with a lattice constant $a = 564$ pm. The crystal was heated to 350$^\circ$C overnight, cleaved inside the load-lock and was subsequently transferred into the preparation chamber, where it was heated to about 300$^\circ$C for 1 h to remove for residual surface charges generated during the cleavage process. An example of a 500 nm $\times$ 500 nm topographic image recorded at 519 mK is depicted in Fig. \ref{fig5}(a). The two line sections in (b) show step heights that correspond to integer multiples of the interlayer distance $a/2$. They also show a time dependent appearance of low frequency noise in the data that is absent in the lower part of the image (black line), but clearly visible in the upper part (red line). The origin of this noise is discussed in the next section. Although we were not able to get rid of the noise, it is possible to obtain atomic resolution at about 550 mK as demonstrated in Fig. \ref{fig6}(a). The line section in (b) shows an atomic corrugation of about 60 pm with a noise level of about $\pm 4$ pm. In addition, we tested the influence of an applied magnetic field on the scanning process. As expected, no negative influence could be observed, as we used non-magnetic materials to build our microscope. In fact, we found a reduction of vibrational noise, probably due to eddy-current damping.

As visible Figs. \ref{fig4}(a) and \ref{fig5}(b) (red line section), excessive low frequency noise (<50 Hz) can be often observed in our image data. We found that these are mechanical vibrations that are generated whenever the insert touches the cryostat tube or the microscope body touches the 1-K-radiation shield. Calculations indicate that the observed frequencies could stem either from eigenmodes of the bellow between 1-K-pot and $^3$He-pot (the thermal switch), from the spiraling capillary, or from bending modes of the hollow insert rod. 

\begin{figure}
\centerline{\includegraphics[width=0.9\linewidth]{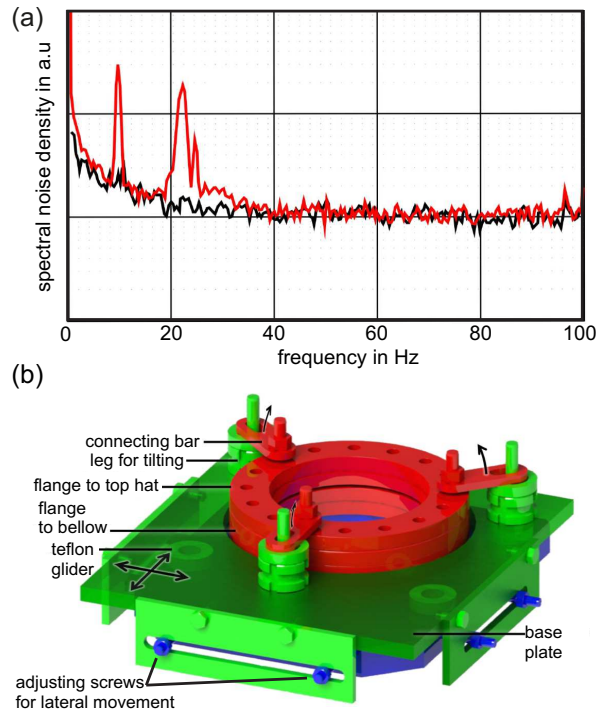}}
\caption{(a) Drawing of the port aligner mounted between below and top hat to adjust the tilt and the lateral position of the insert. (b) Piezoelectric signal from the $\pm x$ electrodes of the sample scanner recorded with a differential amplifier with the insert hanging outside the cryostat. Red and black curves were recorded with the microscope body touching and not touching the  1-K-radiation-shield.}
\label{fig7}
\end{figure}

These vibrations can be minimized by carefully centering the microscope body with respect to the 1-K-radiation shield during assembling using counterweights to compensate for the uneven mass distribution of the microscope body (see Fig. \ref{fig3}). Additionally, a home-built port aligner, cf., Fig. \ref{fig7}(a), mounted between top hat and edge welded bellow (see Fig. \ref{fig2}), can be used to adjust lateral position and tilt of the insert (1-K-radiation shield and radiation baffles) relative to the insert tube. The result of such an aligning procedure is shown in Fig. \ref{fig7}(b). Here, we monitored the vibrations before and after alignment by feeding the signal from two opposing electrodes of the piezoelectric sample scanner into a differential amplifier and recording the data with a spectrum analyzer. Monitoring the static deflection of the cantilever while far away from the sample is also possible.

\section{Conclusion}

We described the set-up of an atomic force microscope with an all-fiber interferometric detection scheme inside of a bakeable $^3$He cryostat equipped with a 10 T magnet that is connected to an ultrahigh-vacuum system. With the AFM in operation a hold time of about 7 h can be achieved at a base temperature of about 550 mK. Currently, the minimal temperature as well as the maximal hold time is limited by the intensity of light required for the interferometric detection, because some fraction of the light is always lost and heats the microscope. The amount of stray light can be strongly reduced by increasing the reflectivity of the cantilever backside as well as of the fiber end, so that both reflectivities are about equal and larger than 90\%. This can be achieved by dielectric mirror coating on the fiber and a metal coating on the cantilever.   
The design allows for an \textit{in-situ} exchange of tip and sample. As demonstrated, \textit{in-situ} preparation of clean metallic as well as insulating substrates is possible. Moreover, single adsorbates can be deposited in a controlled fashion onto cold substrates located inside the microscope. Finally, imaging with atomic resolution is possible at base temperature. The origin of low frequency noise visible in our data is related mechanical vibrations. They can be minimized by proper alignment between microscope body, 1-K-radiation shield, radiation baffles and the inner wall of the cryostat tube. 

\section*{Acknowledgments}

During planning, construction and setting up of the system fruitful discussions with the engineering team from Oxford instruments (particularly R. Vianna), J. Wiebe, R. Schmidt, K. L\"ammle and J. Grenz were very helpful. We are also indebted to the group of Prof. Sengstock for providing the coating for the fiber end. Financial support from SFB 668 and ERC Advanced Grant ASTONISH is gratefully acknowledged.


\begin{references}
\bibitem{MACOR} MACOR is a trademark of Corning Glass Works and denotes a machinable glass ceramic.
\end{references}
\end{document}